# Grease the gears for a steady microfluidic flow

Moritz Leuthner*, and Oliver Hayden*

Heinz-Nixdorf-Chair of Biomedical Electronics, School of Computation, Information and Technology & Munich Institute of Biomedical Engineering, Technical University of Munich, TranslaTUM, Einsteinstraße 25, 81675 Munich, Germany.

Email: moritz.leuthner@tum.de, oliver.hayden@tum.de

Phone: +49 8641 409032

ABSTRACT

Pumps are indispensable for analytical applications and ensure controlled fluid movement. Syringe pumps are among today's most prevalent liquid delivery systems, especially for high-pressure, stable, low-flow-rate microfluidic 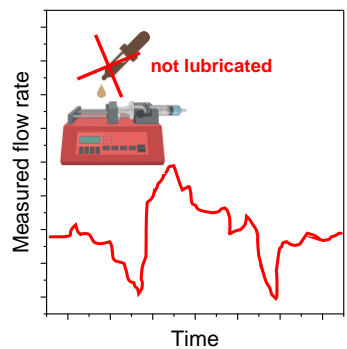 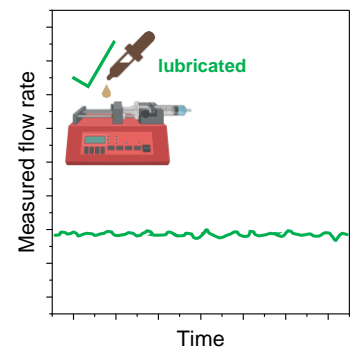 applications. Due to moving mechanical parts of the assembly, regular maintenance is essential to ensure reliable operation and flow rates. However, lubrication of the mechanics is easily overlooked because the research focuses on novel analytical applications rather than on the maintenance of pumps. Here, we investigate the lubrication of the syringe pump guide rods with its effect on the flow rate stability after regular pump cleaning from contaminations. The guide rods of syringe pumps were thoroughly cleaned from any lubricant, and the flow rate for specified flowrates between 5 and 30 µL/min was measured, revealing tremendous flow rate fluctuations with a coefficient of variation (CV) value up to 0.34. In contrast, flow rate measurements of syringe pumps with lubricated guide rods show a five-fold smoother flow rate fluctuation depending on the specified flow rate with CV values below 0.07. In summary, we emphasize the awareness of lubricating moving parts of syringe pumps to achieve constant flow rates, minimize wear, and ensure the reliable operation of, for instance, accurate lab-on-a-chip workflows.

INTRODUCTION

To prepare this article, we asked many microfluidic community colleagues how they ensure their pumps' stable operation. Surprisingly, we learned that most laboratories consider their pumps as plug-and-play tools without an awareness of the maintenance efforts of the mechanics.

Fluid delivery systems are widely used in microfluidic systems with biological, chemical, or medical applications.[1-6] Mainly, syringe pumps, pressure pumps, and peristaltic pumps are employed, besides niche applications for gravity-, osmotic-, surface tension-, electrokinetic-, and centrifugal-driven pumps.[2,7-9] Flow fluctuations can be present in any fluidic system inherited from the fluid driving system, even when static conditions are imposed.

Peristaltic pumps exert a mechanical action on a flexible, liquid-carrying element. The incorporated progressive squeezing action drives the fluid by creating a pressure difference. Here, fluid flow pulsations are unavoidable due to the discontinuous compression of the flexible tubing. Thus, precise dosing is limited, making this type of pump primarily suitable for applications with high flow rates or fluid recirculation.[7]

In contrast, pressure pumps pressurize a fluid-containing reservoir with gas to establish a fluid flow in the microfluidic system. Due to the possibility of exact gas pressure control, very stable flow rates can be achieved in the nL/min-range with even fL-dosing applications.[10,11] However, the achieved flow highly depends on the fluidic resistance of the microfluidic system and, consequently, is directly dependent on the channel geometry and fluid viscosity. This generally excludes its use for applications demanding high pressures and insensitivity to mechanical disturbances.[12]

Today, syringe pumps are usually the most economical solution and, thus, widely used systems for academic laboratories with defined flow rates.[13-16] For instance, a commercial system, such as the Pump 11 Pico Plus Elite from Harvard Apparatus, Holliston, MA, USA, provides a steady flow precision of 1.8±0.08 µL/min with a theoretical minimal flow rate of 2.66 nL/min and sufficient pressure for typical nL-dosing applications.[2,16,17] The motor drives the rotation of a drive screw, translated into a translational motion of the pusher block. Rods or other mechanical structures usually guide the pusher block linearly, picking up the torque. Thereby, the plunger of a mounted syringe is moved by the pusher block resulting in a fluid flow. Mainly, the steps of the motor limit the minimal dispensed volume of a syringe and guarantee a steady flow rate. Li et al. (ref.18) reported that minute fluctuations in the flow rate with frequencies in the 300 – 400 Hz regime could be traced back to the discrete steps taken by the driving stepper motor.[1,18-20] Potentially, other mechanical syringe pump parts can also impede homogeneous fluid dispensing, but to the best of our knowledge, little of their influence is reported. Especially when cleaning a syringe pump thoroughly or just by frequent use, lubricants from moving parts will inevitably be removed and often not sufficiently reapplied. Such cleaning is, for instance, mandatory when removing pumps from a biosafety level 2 environment.

In this article, we demonstrate the importance of lubricating moving parts of a syringe pump to achieve robust analytical results required for precise fluid control. In particular, we compare the fluctuations of flow rates of syringe pumps with and without lubricated guide rods. By exemplifying the forces and torques on the pusher block, we deduce that the canting and stuttering are sources of the fluctuations.

RESULTS AND DISCUSSION

With our setup, we recorded flow rates for lubricated and not-lubricated syringe pump guide rods and compared them regarding flow rate fluctuation during continuous pumping. Additionally, transient phenomena were recorded when repetitively starting and stopping the syringe pump. The dynamic flow rate profiles, quantitative flow rate analysis, and frequency spectra are shown in Figure 1. The steep in- and decrease in measured flow rates reflect the rigid and gas bubble-free fluidic system with no or negligible hydraulic capacitance, allowing a precise evaluation of the syringe pump performance.

**Figure 1.** Comparison of lubricated and not lubricated guide rods regarding flow rate stability.

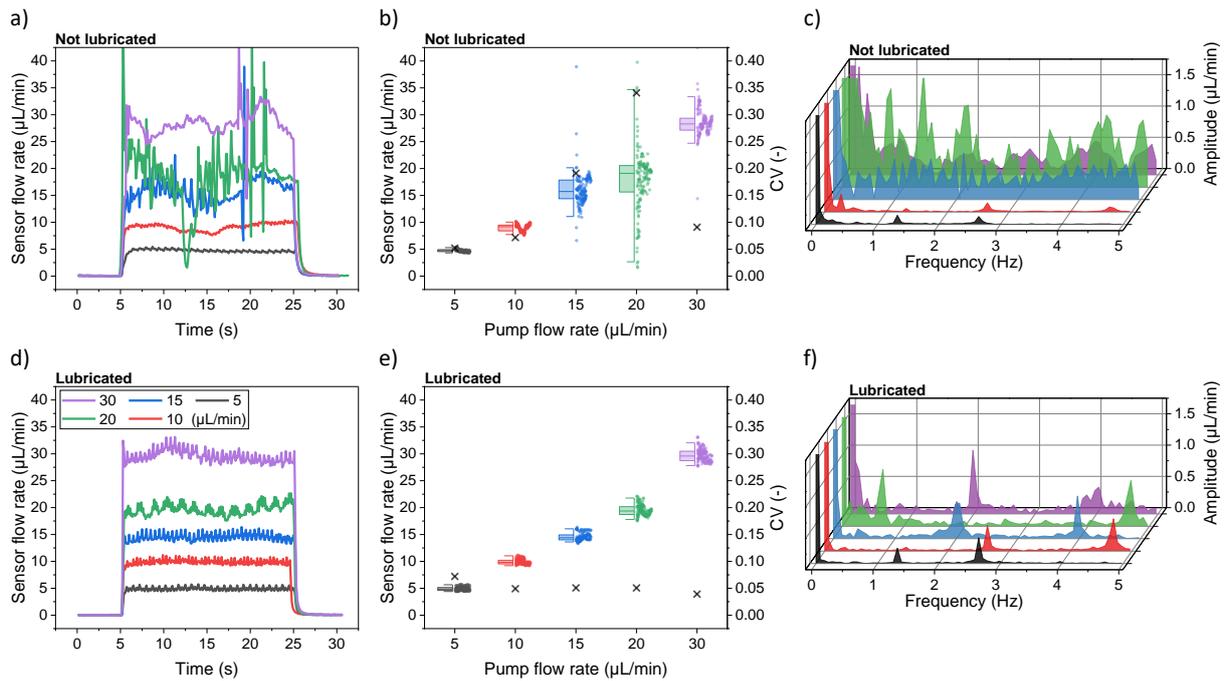

Dynamic flow rate profiles and quantitative analysis of the flow rate stability for lubricated (d), e)) and not lubricated (a), b)) guide rods. The specified, color-coded flow rates range from 5 to 30 µL/min. For plots b) and e), the flow profiles between 7.5 – 22.5 s were evaluated and plotted with whiskers at the 2.5 and 97.5 percentile of the data. The data's coefficient of variation (CV) is plotted with black crosses (x) on the right y-axis. c) and f) show the flow profiles between 7.5 – 22.5 s in the frequency domain by FFT for the not lubricated and lubricated syringe pump, respectively.

*Flow rates without lubricated guide rods.* The specified flow rate was reached immediately after starting the syringe pump. Anomalous fluctuations occurred depending on the flow rate, especially for fixed 15 - 30 µL/min flow rates. At these experimental conditions, striking pulses towards higher and lower flow rates were measured; no steady flow rate condition was achieved during the 20 s pumping interval. Temporary drifts were observed above 10 µL/min. The box plots of the recorded flow rates showed a distribution with broad whiskers at the 2.5 - 97.5 percentiles and CV values ranging from 0.05 to 0.34. Due to torque and high frictional forces, these heterogeneous flow rates must have directly resulted from the pusher block movement that was canting on the guide rods. Visualizing the force from the syringe plunger and the drive screw (Figure 2), resulting moments on the pusher block were induced in y- and z-direction. The magnitude of these moments could be determined by the cross-product of the force and position vector. Due to their point of attack, both forces contributed to the same torque direction in the respective y- and z-direction. Combined with the friction coefficient for brass on steel and some unavoidable play of the fit, canting of the pusher block and unsteady movement were promoted. Hence, reducing the torque of the pusher block or lowering the friction in the fit would redress canting and foster smooth movement. Since geometric modifications of the syringe pump were impossible, the guide rods were lubricated to reduce friction. In contrast, the flow rate at 5 µL/min held a constant level over the whole pumping time but only showed a small saw-tooth-like profile induced by the discrete steps of the pusher block. Only at such low flow rates the CV of the flow rate was 0.05 and comparable to operations with lubricated guide rods. With FFT, the anomalous flow rates can be further quantified with a broad range of frequencies from 15 to 30 µL/min. Only discrete contributions at 1.32 Hz, 2.65 Hz, and 4.70 Hz are observed at 10 µL/min, whereby the two lower frequencies are also present at 5 µL/min.

**Figure 2.** Guiding and driving parts of the syringe pump under test.

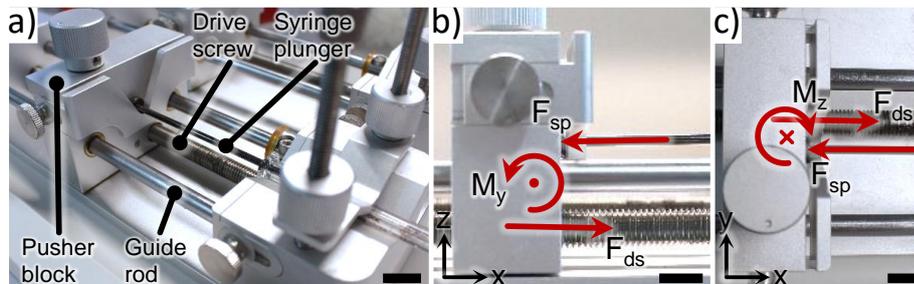

a) Overview of the guide rods and drive screw configuration with mounted syringe. Side view b) and top view c) of the pusher block with acting forces and moments when infusing. $F_{sp}$: force from the syringe plunger; $F_{ds}$: force from the drive screw; M: resulting moment from forces. Scale bars equal 1 cm.

*Flow rates with lubricated guide rods.* All set flow rates showed over the pumping time of 20 s a constant level with small fluctuations around a steady mean. Only at a flow rate of 20 µL/min an additional low-frequency flow rate fluctuation was measured. The impact of the lubrication is clearly observed with CV values around 0.05 for all flow rates but 0.07 for 5 µL/min. Overall, canting of the pusher block was avoided since no disruptive flow rate peaks or significant deviations from the mean flow rate were observed. The lubrication of the guide rods reduced the friction substantially to facilitate the sliding of the pusher block. This is essential for steady flow rates and precise volume dosing with syringe pumps. Moreover, it contributes to less abrasion and minimized pump wear, besides assuring reliable functionality. However, significant components remain in the frequency domain for all flow rates. The 5 – 10 µL/min flow rate peaks are the same as for the measurement without lubricated guide rods but slightly more pronounced (1.32 Hz, 2.65 Hz, 4.70 Hz). Regarding the higher flow rates, most frequency components could be removed except for some remaining peaks. This indicates that not all mechanical disturbances can be eliminated but minimized to some remaining vibrations.

CONCLUSION

Lubrication of mechanical parts is not a novel idea, but it should be mentioned, as the researcher's focus is on the analytics rather than on the maintenance of the periphery. Syringe pumps are mainly used as plug-and-play tools, but for many high-precision applications the noise level of syringe pumps is a significant contributor to low experimental accuracy and may lead to false research results. Here, the flow rate stability of a commercial syringe pump was investigated concerning the lubrication of the moving parts after a typical cleaning cycle. The analysis revealed that in the tested flow regime of 5 – 30 µL/min, the flow rate's CV without lubricated syringe parts could be 680% higher than the lubricated syringe. Striking pulses and unpredictable fluctuations result from the canting pusher block due to insufficient lubrication.

In conclusion, we encourage users of syringe pumps to regularly lubricate all moving parts to improve the experimental quality and reproducibility of microfluidic results. Furthermore, self-built and commercial syringe pumps should undergo a critical design assessment focusing on minimizing torques and play in fits. Despite the widespread syringe pump designs for self-built syringes employing affordable guide rods, it is advisable to implement low-tolerance linear axis, e.g., pre-stressed with dovetail guiding.

MATERIALS AND METHODS

*Syringe pumps.* A syringe pump (Fusion 4000, Chemyx Inc., Stafford, USA) with two parallel stainless steel guide rods and an eccentric drive screw, as shown in Figure 2, was used for all investigations. The minimum flow rate specified by the manufacturer for the used 500 µL syringe is 19.57 nL/min at an accuracy of ±<0.35% and precision of ±<0.05%.[21] The pusher block is aluminum and has brass sliding on the guide rods. In the case of the not-lubricated syringe pump, the guide rods were thoroughly cleaned from any lubricant and dirt with ethanol. However, little lubricant inevitably remained on the interior parts of the fit that could only be reached by completely dismantling the syringe pump. The guide rods were lubricated with non-resinous lubricant (Spezialfett, Mutronic GmbH, Rieden, Germany) for subsequent experiments with the lubricated syringe pump. Operating the syringe pump a few times back and forth over the maximum moving distance distributed the lubricant uniformly on the guide rods and allowed creeping into the fit.

*Setup.* A 500 µL gas-tight glass syringe (1750 TLL, Hamilton Company Corp., Reno, USA) with a 3.26 mm inner diameter, PTFE (Polytetrafluoroethylene) plunger tip, and Luer-lock connection was used to eliminate syringe deformations. A matching dosing tip (Nordson Corp., Westlake, USA) with a stainless steel tip of 0.013" inner and 0.025" outer diameter was used to connect the syringe and tubing. The 12 cm long PFA (perfluoroalkoxy alkane) tubing (IDEX Health & Science LLC, Oak Harbor, USA) with 0.020" inner and 1/16" outer diameter was pushed on the tip forming a tight connection. In contrast, the other end was connected flangeless to a 6 cm long FEP (fluorinated ethylene propylene) tubing (Fluigent S.A., Le Kremlin-Bicêtre, France) with 0.020" inner and 1/32" outer diameter. This thinner tubing was eventually connected to the flow sensor (Flow Unit M, Fluigent S.A.). At the outlet of the flow sensor, the same FEP tubing with a 4 cm length was connected and dipped into a water-pre-filled Eppendorf tube. Please note that the fluidic system had no elastic parts and was realized as rigid as possible to avoid any fluidic capacitance affecting transient phenomena and damping. The flow sensor was connected to its controller (Flow-EZ, Fluigent S.A.), facilitating a readout with a PC.

*Flow Rate Measurements.* The water-filled syringe was fixed tightly in the syringe pump and connected to the fluidic system. Special attention was given to a gas bubble-free filling of the whole system to avoid any fluidic capacitance. 5 s after starting the recording of the flow rate, the syringe pump was started manually. After 20 s of pumping time, the syringe pump is manually stopped, and for another 5 s, the flow rate is recorded to observe transient phenomena, if any. The set flow rates varied between 5 - 30 µL/min. The recording was facilitated with a custom-written PC program (LabVIEW 2018, National Instruments Corp., Austin, USA) and a sample rate of 10 Hz. Quantitative flow rate analysis and the Fast Fourier Transformation (FFT) were conducted in OriginLab Pro 2021b (OriginLab Corp., Northampton, USA). Since we are not interested in flow rate fluctuations induced by the incremental steps of the driving stepper motor but in low-frequency effects, the sample rate of 10 Hz is sufficient for an FFT analysis with a bandwidth of 5 Hz limited by the Nyquist-Shannon sampling theorem. For comparison, with a step resolution $s$ of our syringe pusher block of 0.012 µm/step, we would expect pulses with a frequency between 832 - 4992 Hz determined by $f_{pulse} = 4Q\pi^{-1}D^{-2}s^{-1}$ where $Q$ denotes the set flow rate and $D$ the syringe's inner diameter.[2,21]

ACKNOWLEDGMENTS

This project is supported by the Federal Ministry for Economic Affairs and Climate Action (BMWK) on the basis of a decision by the German Bundestag (Grant No.: KK 5333701AD1).

CONFLICT OF INTERESTS

All authors declare no conflict of interest.

AUTHOR CONTRIBUTIONS

ML and OH designed the experiments and wrote the manuscript. All authors have approved the final version of the manuscript.